\documentclass[%
 reprint,
superscriptaddress,
 amsmath,amssymb,
 aps,prl,nofootinbib
]{revtex4-1}

\usepackage{graphicx, epsfig, array, subfigure, epstopdf, xcolor, adjustbox, soul}
\usepackage[colorlinks=true, citecolor=blue, linkcolor=blue, urlcolor=blue]{hyperref}

\newcommand{\mA}{m_{\gamma^\prime}}

\begin{document}

\title{New Exclusion Limit for Dark Photons from an SRF Cavity-Based Search (Dark SRF)}
\author{A. Romanenko} 
\email{aroman@fnal.gov}
\author{R. Harnik}
\email{roni@fnal.gov}
\author{A. Grassellino}
\email{annag@fnal.gov}
\author{R. Pilipenko}
\author{Y. Pischalnikov}
\affiliation{Fermi National Accelerator Laboratory, Batavia, IL 60510, USA}
\author{Z. Liu}
\email{zliuphys@umn.edu}
\affiliation{School of Physics and Astronomy, University of Minnesota, Minneapolis, MN 55455, USA}
\author{O. S. Melnychuk}
\author{B. Giaccone}
\author{O. Pronitchev}
\author{T. Khabiboulline}
\author{D. Frolov}
\author{S. Posen}
\affiliation{Fermi National Accelerator Laboratory, Batavia, IL 60510, USA}
\author{A. Berlin}
\affiliation{Fermi National Accelerator Laboratory, Batavia, IL 60510, USA}
\author{A. Hook}
\affiliation{Maryland Center for Fundamental Physics, University of Maryland, College Park, MD 20742}


\date{\today}

\begin{abstract}
We conduct the first ``light-shining-through-wall" (LSW) search for dark photons using two state-of-the-art high quality-factor superconducting radio frequency (SRF) cavities and report the results of its pathfinder run. 
Our new experimental setup enables improvements in sensitivity over previous searches and covers new dark photon parameter space. 
We design delicate calibration and measurement protocols to utilize the high-$Q$ setup at Dark SRF.
Using cavities operating at $1.3 \ \text{GHz}$, we establish a new exclusion limit for kinetic mixing as small as {$\epsilon= 1.6\times 10^{-9}$} and provide the world's best constraints on dark photons in the $2.1\times 10^{-7} \ \text{eV} - 5.7\times10^{-6} \ \text{eV}$ mass range.
Our result is the first proof-of-concept for the enabling role of SRF cavities in LSW setups, with ample opportunities for further improvements. 
In addition, our data sets a competitive lab-based limit on the Standard Model photon mass by searching for longitudinal photon polarization. 
\end{abstract}


\maketitle

\section{Introduction}
One of the conceptually simplest proposed extensions to the Standard Model (SM) of particle physics is the existence of a hidden sector consisting of particles feebly coupled to ordinary matter. In particular, the hypothesized ``dark photon''~\cite{Holdom:1985ag} interacts with ordinary matter via a small kinetic mixing with the SM photon. 
A range of possible photon-dark photon couplings $\epsilon$ and dark photon masses $m_\mathrm{\gamma'}$ has been previously excluded based on the combination of laboratory experiments and astrophysical observations~\cite{Jaeckel:2010ni,Caputo:2021eaa}. 

Several lab-based dark photon searches have been performed using the ``light-shining-through-wall" (LSW) scheme~\cite{Okun:1982xi,VanBibber:1987rq,Hoogeveen:1990vq}. These experiments involve a source of SM photons at a particular frequency (e.g., laser light), a wall impenetrable to this light, and a detector looking for photons of the same frequency that emerge past the wall. The emission of some of the photons as dark photons before the wall and their detection after the wall makes such a search possible if dark photons exist with a hypothesized mass and coupling. Resonant cavities can be used on both sides of the wall to increase the number of photons on the emitting side and to enhance the detection probability 
on the receiver side. This scheme has been implemented successfully in the optical~\cite{Ehret:2009sq} and microwave frequency regimes~\cite{Betz:2013dza, Parker:2013fxa}. 

The use of superconducting microwave cavities in LSW experiments was proposed in Ref.~\cite{Jaeckel_PhysLettB_2008}, and an optimal relative cavity orientation to emit and detect the longitudinal dark photon polarization was later identified in Ref.~\cite{Graham_PRD_2014}. Compared to optical light, microwaves enable the employment of simpler cavity engineering and readout and state-of-the-art ultra-high quality factor cavities. The most recent LSW experiments in the microwave regime utilized two  normal-conducting cavities with loaded quality factors of $Q\sim10^3-10^4$~\cite{Betz:2013dza, Parker:2013fxa}. Superconducting radio frequency (SRF) cavities routinely utilized in modern particle accelerators~\cite{Padamsee_Ann_Rev_Nucl_2014} have intrinsic quality factors of $Q>10^{10}$, providing a unique opportunity for multiple orders of magnitude enhancement both in the number of stored photons in the ``emitter'' cavity and in the detection sensitivity of the ``receiver'' cavity. In this work, we present the first results of Dark SRF, an LSW experiment utilizing such ultra-high quality SRF cavities with an optimal arrangement for longitudinal dark photon detection.

\section{The Dark Photon and Signal Power}

The dark photon $A'_\mu$ is a massive vector boson of mass $\mA$ which interacts with the SM via
\begin{equation}
    \mathcal{L} = \mathcal{L}_\mathrm{SM} - \frac{1}{4} \, F'_{\mu\nu}F'^{\mu\nu} + \frac{\epsilon}{2} \,  F'_{\mu\nu} \, F^{\mu\nu} +\frac{1}{2}\, \mA^2 \, A'_\mu A'^{\mu}
    ~,
\end{equation}
where $F$ and $F'$ are the field strengths of the photon and dark photon, and $\epsilon$ is a small kinetic mixing parameter which couples the SM and dark electromagnetic fields. As a result, a coherently oscillating SM electromagnetic field acts as a source of dark photons at the same frequency. In addition, a coherent dark photon field can resonantly excite an RF cavity with the appropriate frequency. 

Given an emitter cavity with intrinsic quality factor $Q_\mathrm{em}$ and stored energy $U_\mathrm{em}$, the radiated dark photon field will then deposit power in a nearby receiver cavity with intrinsic quality factor $Q_\mathrm{rec}$, 

\begin{equation}
    P_\mathrm{rec}=\epsilon^4 \left(\frac{\mA}{\omega}\right)^4 \left|G\right|^2 \, \omega\, Q_\mathrm{rec} \,  U_{\mathrm{em}}
    \label{eq:sig}
    ~,
\end{equation}

where $G$ is a form factor specified in the appendix, which heuristically consists of the wavefunction overlap of the dark photon field with the spatial mode shape in the receiver cavity.
As pointed out in Ref.~\cite{Graham_PRD_2014}, when the cavities are oriented to produce and detect the longitudinal polarization of the dark photon field, the right-hand-side of Eq.~(\ref{eq:sig}) is suppressed by the fourth power of mass over frequency rather than the eighth power, as was the case in the dark photon search in the CROWS experiment~\cite{Betz:2013dza, Parker:2013fxa}, which was optimized for the transverse mode.\footnote{As was also pointed out in Ref.~\cite{Graham_PRD_2014}, the axion search conducted by CROWS was arranged longitudinally and may be reinterpreted to give a stronger dark photon limit at low masses.} 
Assuming that the noise in the receiver is thermal, the signal-to-noise ratio (SNR) is given by the radiometer formula~\cite{Dicke:1946glx}
\begin{equation}
    \mathrm{SNR}=\frac{P_\mathrm{rec}}{P_\mathrm{th}} \, 
    \sqrt{\delta\nu\, t_\mathrm{int}} =
    \frac{P_\mathrm{rec}}{k_B T_\mathrm{eff}} \, 
    \sqrt{\frac{t_\mathrm{int}}{\delta\nu}}
    ~ ,
    \label{eq:SNR}
\end{equation}
where $\delta\nu$ is the bandwidth of the analysis, $t_\mathrm{int}$ is the integration time, $T_\mathrm{eff}$ is the effective noise, and  $P_\text{th} = k_B \, T_\text{eff} \, \delta \nu$ is the noise power. The factor of $\sqrt{\delta\nu\, t_\mathrm{int}}$ in the first equality is the square root of the number of independent measurements. 

\section{Experimental setup}
\begin{figure}[htb]
 \includegraphics[scale=0.7]{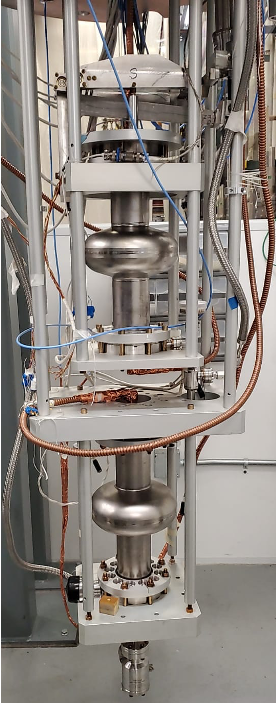}\qquad 
 \includegraphics[scale=0.36]{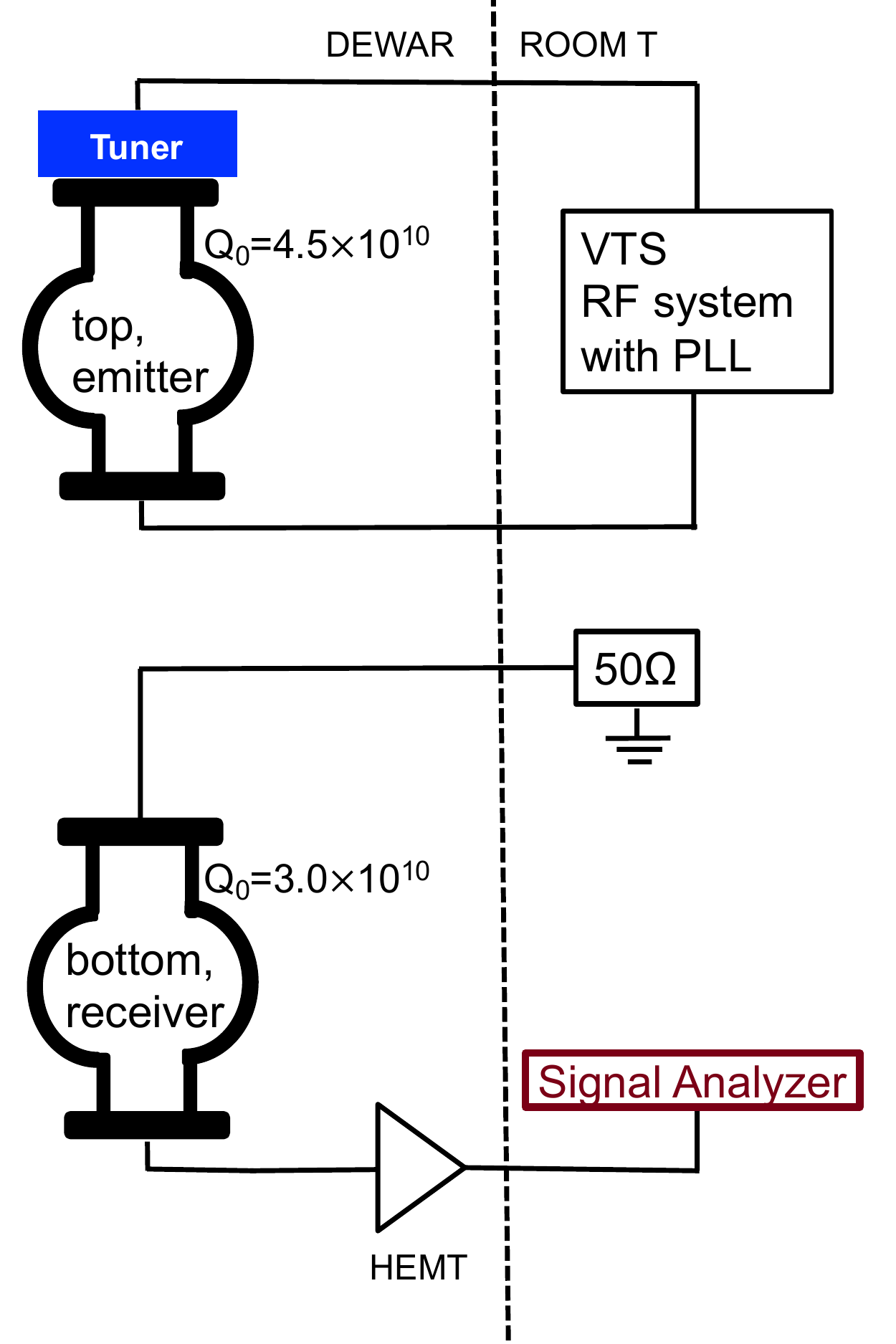}
\caption{\label{fig:DarkSRFpic}Left: The experimental setup for the Dark SRF experiment consisting of two 1.3~GHz cavities. Right: A sketch of the Dark SRF electronic system.}
\end{figure}

The experimental setup, shown on the left of Fig.~\ref{fig:DarkSRFpic}, has been assembled using two 1.3~GHz high-quality factor $Q_0$ SRF cavities. Both cavities are made out of bulk high residual resistivity ratio $\text{(RRR)}> 200$ niobium and have been prepared by bulk electropolishing, 800$^\circ$C annealing for 3 hours, light electropolishing, and a final 120$^\circ$C 48 hours baking. The cavities have been selected to have resonant frequencies of the fundamental TM$_{010}$ modes as close to each other as possible. The mechanical holding structure has been designed in such a way as to minimize the frequency shifts of the receiver (bottom) cavity in response to liquid helium pressure fluctuations and mechanical vibrations, while allowing the emitter (top) cavity to be frequency-tunable by tiny mechanical deformations. The cavity centers were mounted 60~cm apart (center-to-center), and the cavities were oriented along a common axis. An accelerator-style frequency tuner~\cite{Pischalnikov_IPAC2015} assembly has been attached to the emitter cavity, allowing both ``coarse'' tuning by a stepper motor in a range of about 5~MHz with $\sim 12 \ \text{Hz}$ resolution, as well as ``fine'' tuning using the piezoelectric element in a range of about 8~kHz with $\sim 0.1 \ \text{Hz}$ resolution~\cite{Pischalnikov_SRF2019}. 
Shielded microwave coaxial cables independently connect each cavity to room temperature signal generation and measurement electronics. 

Initially, both of the cavities were tested at $2 \ \text{K}$ and $1.5 \ \text{K}$ in a liquid helium bath, using standard SRF techniques~\cite{Melnychuk_RSI_2014} to determine the fundamental mode frequencies and intrinsic quality factors $Q_0$, as well as the input ($Q_\mathrm{in}$) and transmitted ($Q_\mathrm{t}$) antenna external quality factors. The emitter cavity RF ports were connected directly to incident power and transmitted power cables of the RF system in Fermilab's Vertical Test Stand (VTS) and the corresponding phase lock loop (PLL) system. In frequency stability runs, the receiver cavity input RF port was connected to $30 \ \text{dB}$ attenuator to suppress room temperature thermal noise leaking into the cavity. The receiver cavity output RF port was connected to a high electron mobility transistor (HEMT) amplifier (LNF-LNC0.3\_14A) to amplify the dark photon signal. The HEMT amplifier had a nominal gain of $38 \ \text{dB}$ at $5 \ \text{K}$. The HEMT nominal noise temperature at $5 \ \text{K}$ was $4 \ \text{K}$. From thermal noise measurements at $1.3 \ \text{K}$ (no RF power delivered to cavities) we estimated an effective gain of the HEMT, including attenuation of the connecting cables, to be $37 \ \text{dB}$, corresponding to a noise temperature of $6.2 \ \text{K}$.

\subsection*{Frequency Stability}

\begin{figure}
    \centering
    \includegraphics[width=8.65cm]{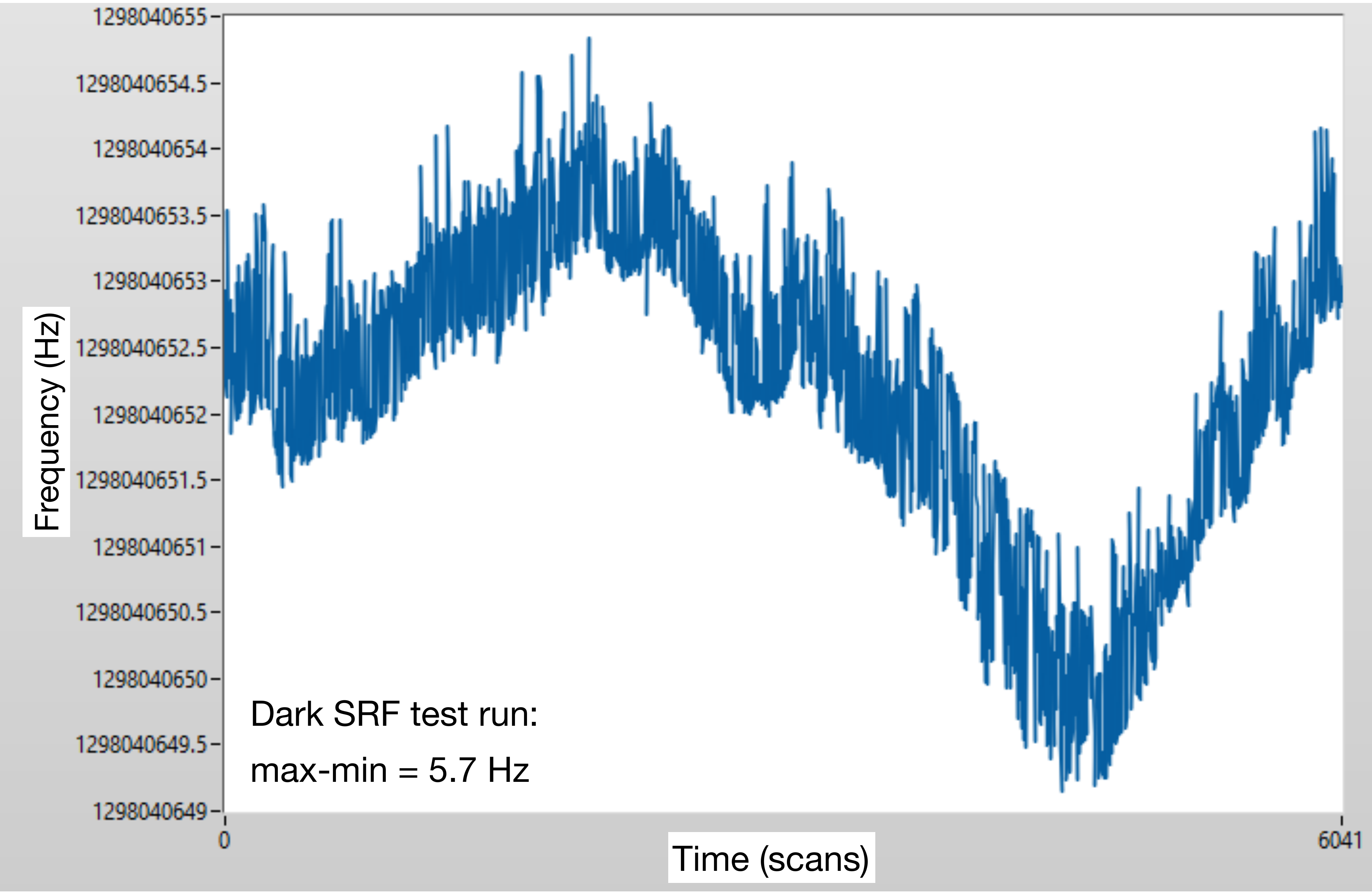}
    \caption{The Dark SRF emitter frequency collected over 6041 scans (each lasting about a second). The frequency variation in this test spans 5.7~Hz. The emitter cavity was the less stable of the emitter-receiver pair.}
    \label{fig:freq}
\end{figure}

Frequency instability of the cavities may be separated into a slow drift on the time scale of minutes or more rapid variations known as microphonics. The frequency drift of the Dark SRF cavities was characterized by frequency stability runs during which the emitter cavity was driven on resonance with a PLL, and the resonant frequency was measured using a frequency counter. 
In a dozen frequency stability runs with $E_\text{acc}=6.4 \ \text{MV}/\text{m}$ (corresponding to  $U_\mathrm{em} = 0.6\ \text{J}$), minimum-to-maximum emitter frequency variations between $3 \ \text{Hz}$ and $8 \ \text{Hz}$ were observed at $2 \ \text{K}$. 
No systematic difference was observed between the runs in which the frequencies were matched with the slow tuner and those with a piezo. A frequency stability test conducted within a day of the dark photon search showed frequency variations spanning $5.7 \ \text{Hz}$ over 100 minutes at $1.4 \ \text{K}$, as shown in Fig.~\ref{fig:freq}. 
Given the proximity of this measurement to the dark photon search, we take $5.7 \ \text{Hz}$ as a conservative input for the emitter drift in the analysis below. 
In addition, a special frequency stability run was recorded, in which only the receiver cavity was powered up and phase-locked while the emitter was not powered at all. The field in the receiver was $E_\text{acc}=14 \ \text{kV}/\text{m}$, and the duration of the run was about forty minutes. Minimum-to-maximum frequency variation in this run was 3\,Hz.

The measurement of microphonics was presented in Ref.~\cite{Pischalnikov:2019iyu}. The cavity frequency was measured rapidly as the cavity was being excited in resonance with a PLL in VTS. The RMS of frequency variations was measured to be~3\,Hz. The characteristic timescale for these frequency jitters was $\sim 20-30 \  \text{milliseconds}$. We take $3 \ \text{Hz}$ as the magnitude of microphonics in the analysis below. This frequency stability, along with the slow drift in cavity frequency, affects the experimental sensitivity, as discussed below.

\section*{Data taking}

Various system configurations were used to search for a dark photon. In each run, several steps were followed: 1) frequency matching, 2) dark photon search, 3) frequency re-check, 4) cross-talk check, and 5) thermal background measurement (if no cross-talk was seen). We discuss each step below.

For frequency matching, both cavities have been powered up by a single RF generator signal split into two routes. A phase lock loop has been used to have the RF generator follow the frequency of the emitter cavity, and the tuner mechanism has been applied to change the emitter cavity frequency until both emitter and receiver cavities are resonantly excited by the same signal, as observed with the signal analyzer connected to the receiver transmitted power line. Since the radiation pressure on the cavity walls changes the cavity frequency, as do small variations in the conditions in VTS, this procedure for frequency matching has been performed for each stored energy in the emitter cavity for the dark photon searches.

After frequency matching, the cable connecting the generator to the receiver cavity was disconnected, the receiver input line was terminated, and the measurement of the transmitted power from the receiver cavity was performed using a spectrum analyzer centered around the resonance frequency. The frequency sweeps in a window of $3 \ \text{kHz}$ with a resolution bandwidth of $1 \ \text{Hz}$ were utilized. Each scan used 10,000 points and took 1.83 seconds. Each data acquisition run lasted for $\sim 30 \ \text{minutes}$. This work presents the result of a $E_\text{acc}=6.2\ \text{MV}/\text{m}$ (corresponding to $U_\mathrm{em}=0.6 \ \text{J}$ of stored energy) run using the linear average of $\sim 1300$ frequency sweeps, in which no cross-talk noise was observed.

Upon conclusion of each dark photon search run, the physical cable connection between the generator and the receiver cavity input was reestablished to verify that the cavities remained frequency matched. 

We have performed several additional measurements to evaluate the amount of cross-talk present in the system in each run. Firstly, the phase lock was disengaged to allow the generator and the cavity to lose synchronization and the stored power in the emitted cavity to drop. If a peak of excess power seen in the receiver cavity moves to follow the frequency of the generator, the excess is deemed to be due to cross-talk. Cross-talk was suppressed after efforts to augment shielding in the signal lines, but it was still observed in some configurations. Shielding efforts continue; however, here, we focus on one configuration in which cross-talk was not observed. In the case when cross-talk was not observed, the power to the emitter was turned off to measure the thermal noise in the receiver cavity. 

\begin{figure}[t]
 \includegraphics[width=1\linewidth]{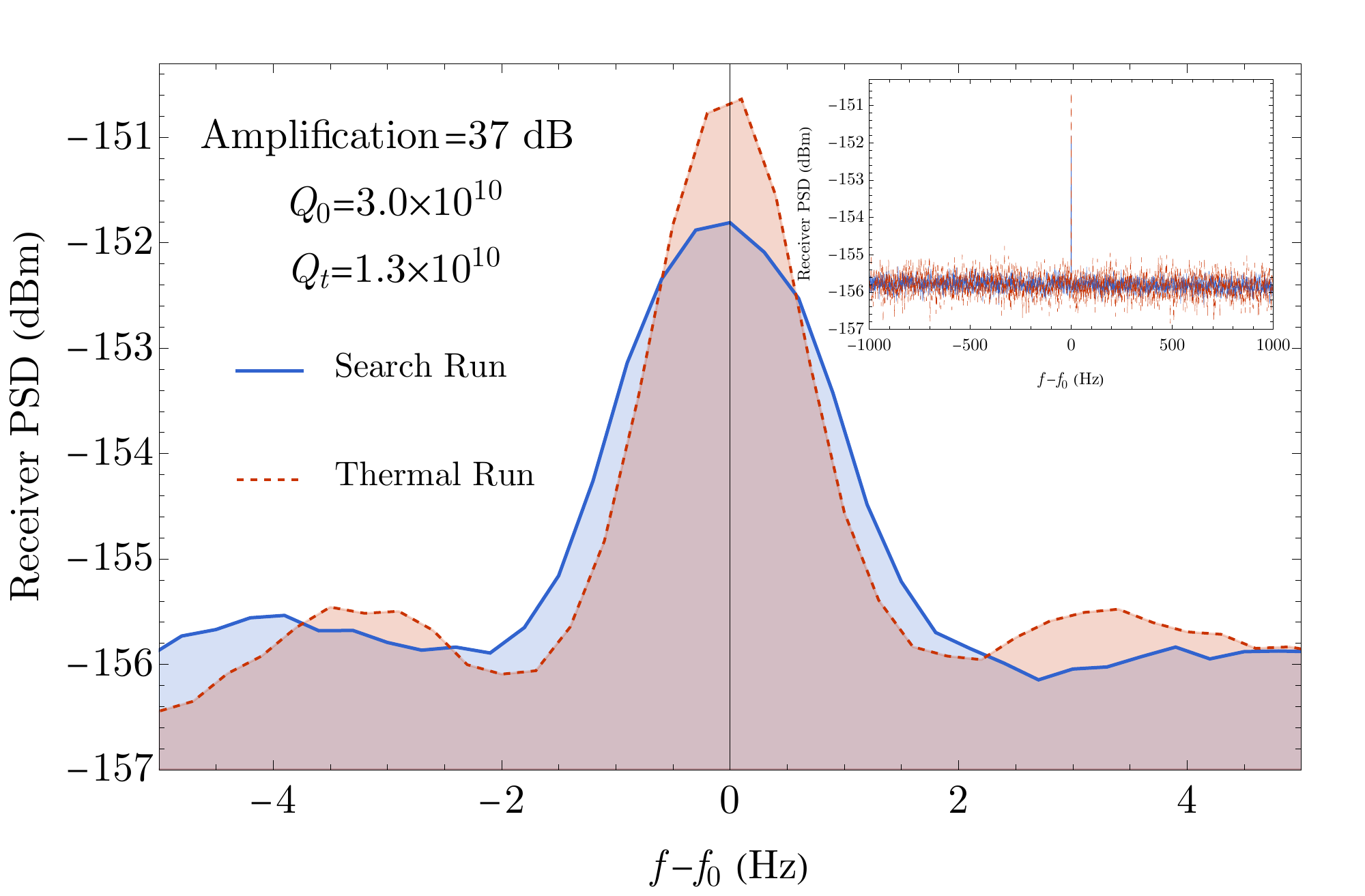}
\caption{\label{fig:data} The measured power spectral density (PSD) of the receiver cavity during the dark photon search run (blue) and thermal noise calibration run (red). For both runs, the peaks are caused by a leak of thermal photons from the receiver input line. The reference frequency is $f_0=1.298041$~GHz. The two peaks were shifted within the drifting modeling by 3.5~Hz to coincide. The signal region is $\pm 1.5 \ \text{Hz}$ around the peak. The power measured between the thermal run ($-151.6^{+0.23}_{-0.25} \ \text{dBm}$) and search run ($-151.8^{+0.16}_{-0.17} \ \text{dBm}$) is consistent with each other within one sigma.
}
\end{figure}

Overall, several data acquisition runs were performed with the emitter field level in the range\footnote{We use $E_\mathrm{acc}$, the effective amount of energy given to a charged particle traversing the cavity, to facilitate the comparison with state-of-the-art SRF accelerator cavities. For the mode in question, this field is related to the peak field by $E_\mathrm{peak}\simeq 1.8 \, E_\mathrm{acc}$.} $E_\mathrm{acc}=6-25 \ \text{MV}/\text{m}$, equivalent to $U_\mathrm{em}=0.6-9.8 \ \text{J}$ of stored energy. We will present the results of one of the low power runs with $E_\mathrm{acc}=6.2$~MV/m, $U_\mathrm{em} = 0.6 \ \text{J}$. Runs with higher stored fields exhibited a larger frequency drift and/or a significant amount of cross-talk. 

\subsection*{Noise analysis}

Fig.~\ref{fig:data} shows spectra measured in both a thermal run (no excitation in the emitter cavity) and a search run. In both cases, a relatively flat power spectrum is observed with a peak centered at the frequency of the receiver cavity. This is the expected signal measured from a HEMT amplifier connected to a cavity that emits thermal photons around its resonant frequency. The peak was shown not to originate from cross-talk, and it also remained when the emitter power was turned off, signifying that it is background rather than signal. 

In the $E_{\rm acc}=6.2\ \text{MV}/\text{m}$ run, the emitter cavity was excited using a $1 \ \text{W}$ amplifier, achieving $0.6 \ \text{J}$ of stored energy. During the dark photon search, the receiver spectrum was well described by a flat background at $-156 \ \text{dBm}$ of transmitted power $P_{\mathrm{t}}$ and a sharp peak of $P_{\mathrm{t}}=-152 \ \text{dBm}$ at the cavity frequency, as shown in Fig.~\ref{fig:data}. These values correspond to the power transmitted by the receiver cavity and include the $37 \ \text{dB}$ of amplification added by the HEMT on the $P_{\mathrm{t}}$ line. The flat background is consistent with the thermal noise in the HEMT amplifier, both in its power level and its variance (sampling over many frequencies). 

The origin of the peak was identified to be a leak of thermal photons from the receiver input line (used to align the two cavities). This hypothesis was corroborated in a subsequent run in which the peak was removed by adding a $30 \ \text{dB}$ attenuator on the input line.\footnote{The runs with this lower thermal background encountered a larger cross-talk peak and were thus not used to set the strongest limit.}

\section{Results}

\begin{figure}[t]
 \includegraphics[width=1 \linewidth]{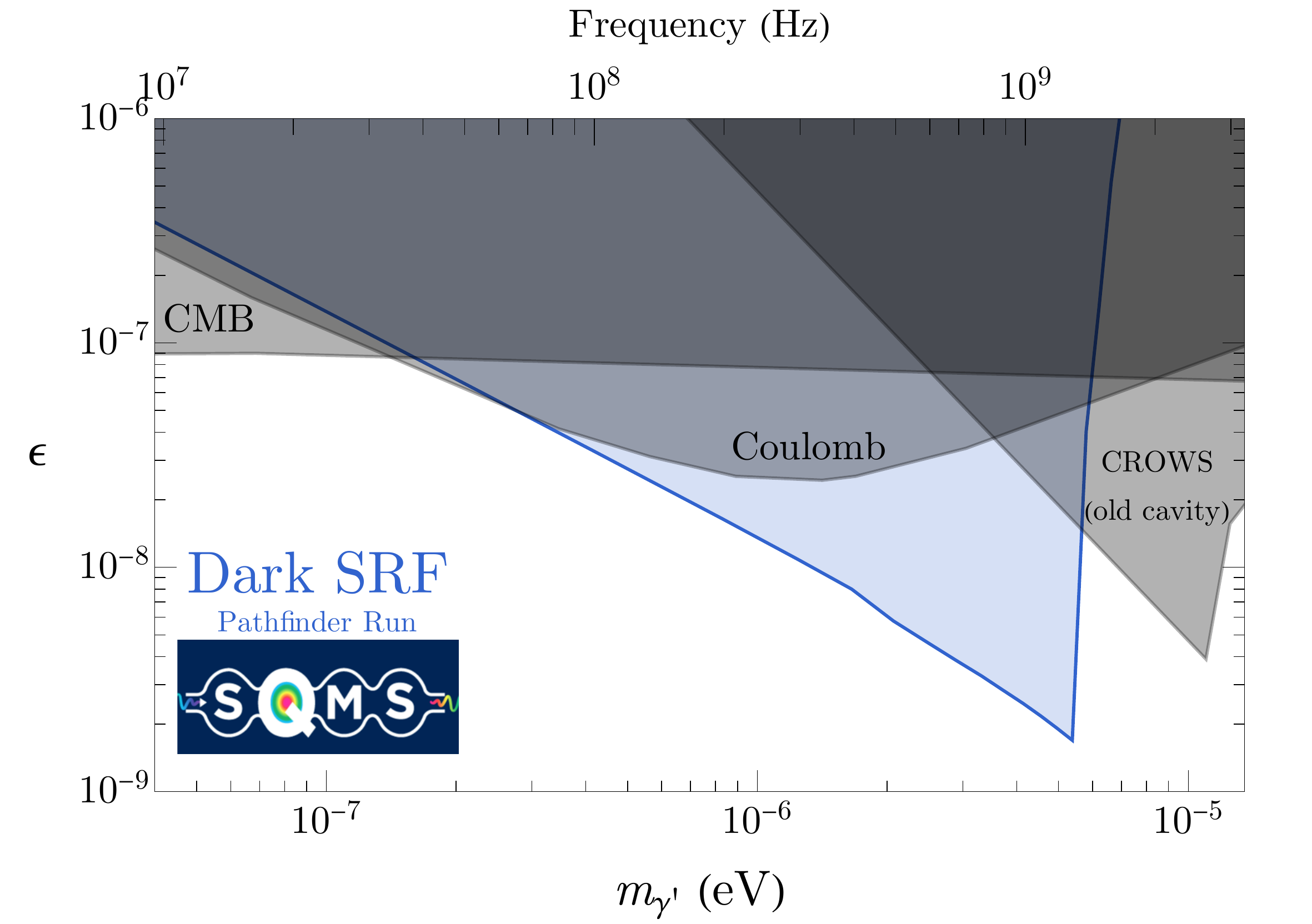}
\caption{\label{fig:Exclusion} The new 95\% C.L. exclusion limit on dark photon parameter space. Our result is shown as the blue curve, where the region above is excluded. 
Also shown in gray are existing limits from the CROWS cavity experiment~\cite{Betz:2013dza}, measurements of the CMB~\cite{Mirizzi:2009iz,Caputo:2020bdy,Garcia:2020qrp}, and tests of Coulomb's law~\cite{Williams:1971ms,Abel:2008ai}.
}
\end{figure}

In the absence of a signal, this run was used to set a leading limit on the dark photon over a wide range of masses, as shown in Fig.~\ref{fig:Exclusion}. The limit setting procedure is described in the appendix with the experimental parameter set shown in Table~\ref{tab:unctobeprop}. The receiver power, accounting for the power in the $\pm 1.5 \ \text{Hz}$ of the peak frequency after considering the amplification and quality factors in different subsystems, is $-189.8^{+0.16}_{-0.17} \ \text{dBm}$, while from the thermal run we know the thermal power around the peak is $-189.6^{+0.23}_{-0.24} \ \text{dBm}$. This allows us to put a limit on the signal power arising from a dark photon for a broad range of masses, predicted using Eq.~(\ref{eq:sig}). In particular, we demand that the signal power not exceed $-189.5 \ \text{dBm}$ (not adding more than $0.3 \ \text{dB}$ beyond the measured $-189.6 \ \text{dBm}$ during the dark photon search run) at 95\% confidence level, using the CLs method~\cite{ParticleDataGroup:2022pth} and Eq.~(\ref{eq:SNR}). 

\begin{table}
\begin{center}
\begin{tabular}{ccc}
\hline\hline
Parameter & \ Emitter \  & \ Receiver \  \\ \hline 
 $Q_0$ & $4.5\times10^{10}$ & $3.0\times10^{10}$ \\
 $Q_\mathrm{in}$ & $1.8\times10^{9}$  & $4.5\times10^{11}$ \\
 $Q_\mathrm{t}$ &  $2.9\times10^{11}$  & $1.3\times10^{10}$ \\
freq. drift & 5.7 Hz & 3.0 Hz \\
microphonics & 3.1 Hz &  3.1 Hz  \\
$P_{\mathrm{loss}}$   &  20 dBm  &  -189 dBm     \\
$U$ & $6.7\times 10^{23}$ & $3.5\times 10^3$\\
\hline\hline
\end{tabular}
\end{center}
\caption[]{Table of key experimental parameters of the Dark SRF low power run used to set dark photon limits. Quality factors (intrinsic $Q_{\mathrm{0}}$ and externals $Q_{\mathrm{in}}$, $Q_{\mathrm{t}}$) reported in the table are known within 10\%. $U$ for the emitter and receiver are the stored number of photons (equivalently, stored energy) in the equilibrium state of the cavities. $P_{\mathrm{loss}}$ is the power lost on the cavity walls, defined as $P_{\mathrm{injected}}-P_{\mathrm{transmitted}}$ or $P_{\mathrm{forward}}-P_{\mathrm{reflected}}-P_{\mathrm{transmitted}}$.
}\label{tab:unctobeprop}
\end{table}

The result is shown in Fig.~\ref{fig:Exclusion}. Our experiment excludes regions above the solid blue line. Compared to existing constraints from the cavity LSW experiment CROWS~\cite{Betz:2013dza}, measurements of the CMB~\cite{Mirizzi:2009iz,Caputo:2020bdy,Garcia:2020qrp} (note, however, that these bounds are slightly alleviated in various models of dark sectors~\cite{Berlin:2022hmt}), and tests of Coulomb's law~\cite{Williams:1971ms,Abel:2008ai}, 
our search provides the world's best limit and improves constraints on $\epsilon$ throughout the dark photon mass range of $\mA \simeq 2.1\times10^{-7} \ \text{eV} - 5.7\times10^{-6} \ \text{eV}$. 
We note that the improvement in $\epsilon$ sensitivity scales as ${\rm SNR}^{\mathrm 1/4}$, clearly showing that our setup is advantageous in many aspects. We stress, however, that this result is a first pathfinder experiment and there are several opportunities for improvement in forthcoming iterations, as discussed in the conclusions.

We also note that this result, as well as any experiments sensitive to the longitudinal mode of the dark photon, can also be interpreted as a limit on the SM photon mass $m_\gamma$. This is because the dark photon longitudinal mode couples directly to SM charges, analogous to the longitudinal mode of a massive SM photon, but with a coupling suppressed by $\epsilon$. In particular, for Dark SRF the formalism describing the production in the emitter cavity, screening in conducting shields, and signal generation in the receiver cavity are nearly identical between the two model scenarios if one equates $\epsilon \, \mA \simeq m_\gamma$ for $\epsilon \ll 1$. Thus, we find that the search described here provides a competitive direct laboratory limit on the SM photon mass~\cite{Workman:2022ynf},  bounding it to be smaller than $8.6 \times 10^{-15} \ \text{eV} \simeq 1.5 \times 10^{-47} \ \text{g}$. It has also been shown in Ref.~\cite{Berlin:2020pey} that this same setup can be sensitive to ultralight ($\ll 1 \ \text{meV}$) ``millicharged" particles that are produced by the large electric fields of the emitter cavity. A simple estimate suggests that the run described here is sensitive to such particles with effective charges as small as $\sim \text{few} \times 10^{-9}$, roughly two orders of magnitude smaller than the best-existing laboratory bounds~\cite{Ahlers:2006iz, DellaValle:2014xoa}. We will perform a dedicated analysis along these lines in future work.

\section{CONCLUSIONS}

We present the first proof of concept experiment of an LSW experiment based on superconducting cavities.
Our experiment was constructed based on high-quality factor SRF niobium cavities, and utilizing accelerator technology for high-precision frequency tuning allowed us to extend the exclusion boundary for the existence of dark photons in a broad range of rest masses and coupling constants.

Future improvements beyond this proof-of-concept run will be pursued. 
The utilization of ultra-high-$Q$ cavities introduces new and unique challenges in frequency stability and control. 
Bringing the frequency stability from a few Hz to the sub-Hz regime will significantly enhance the sensitivity. 
As frequency stability improves even further, beyond-state-of-the-art cavity coherence will lead to deeper sensitivity.
In addition, further suppression of the cross-talk at higher emitter cavity powers can lead to higher signal powers. 
Placing the receiver cavity at milli-Kelvin temperatures, inside a dilution refrigerator and coupled with a quantum-limited amplifier, will lead to lower thermal noise. 
Implementation of phase-sensitive readout can lead to an improved scaling of the SNR with integration time. The combination of these may lead to several more orders of magnitude of explored parameter space~\cite{Berlin:2022hfx}.

\acknowledgments
The authors would like to acknowledge help with measurements from Alexander Netepenko and help with numerical form factor calculations from Josh~Isaacson and Will~Jay.
This material is based upon work supported by the U.S. Department of Energy, Office of Science, National Quantum Information Science Research Centers, Superconducting Quantum Materials and Systems Center (SQMS) under contract number DE-AC02-07CH11359

\bibliography{ROMANENKO_bibliography.bib}

\begin{thebibliography}{29}%
\makeatletter
\providecommand \@ifxundefined [1]{%
 \@ifx{#1\undefined}
}%
\providecommand \@ifnum [1]{%
 \ifnum #1\expandafter \@firstoftwo
 \else \expandafter \@secondoftwo
 \fi
}%
\providecommand \@ifx [1]{%
 \ifx #1\expandafter \@firstoftwo
 \else \expandafter \@secondoftwo
 \fi
}%
\providecommand \natexlab [1]{#1}%
\providecommand \enquote  [1]{``#1''}%
\providecommand \bibnamefont  [1]{#1}%
\providecommand \bibfnamefont [1]{#1}%
\providecommand \citenamefont [1]{#1}%
\providecommand \href@noop [0]{\@secondoftwo}%
\providecommand \href [0]{\begingroup \@sanitize@url \@href}%
\providecommand \@href[1]{\@@startlink{#1}\@@href}%
\providecommand \@@href[1]{\endgroup#1\@@endlink}%
\providecommand \@sanitize@url [0]{\catcode `\\12\catcode `\$12\catcode
  `\&12\catcode `\#12\catcode `\^12\catcode `\_12\catcode `\%12\relax}%
\providecommand \@@startlink[1]{}%
\providecommand \@@endlink[0]{}%
\providecommand \url  [0]{\begingroup\@sanitize@url \@url }%
\providecommand \@url [1]{\endgroup\@href {#1}{\urlprefix }}%
\providecommand \urlprefix  [0]{URL }%
\providecommand \Eprint [0]{\href }%
\providecommand \doibase [0]{http://dx.doi.org/}%
\providecommand \selectlanguage [0]{\@gobble}%
\providecommand \bibinfo  [0]{\@secondoftwo}%
\providecommand \bibfield  [0]{\@secondoftwo}%
\providecommand \translation [1]{[#1]}%
\providecommand \BibitemOpen [0]{}%
\providecommand \bibitemStop [0]{}%
\providecommand \bibitemNoStop [0]{.\EOS\space}%
\providecommand \EOS [0]{\spacefactor3000\relax}%
\providecommand \BibitemShut  [1]{\csname bibitem#1\endcsname}%
\let\auto@bib@innerbib\@empty
\bibitem [{\citenamefont {Holdom}(1986)}]{Holdom:1985ag}%
  \BibitemOpen
  \bibfield  {author} {\bibinfo {author} {\bibfnamefont {B.}~\bibnamefont
  {Holdom}},\ }\href {\doibase 10.1016/0370-2693(86)91377-8} {\bibfield
  {journal} {\bibinfo  {journal} {Phys. Lett. B}\ }\textbf {\bibinfo {volume}
  {166}},\ \bibinfo {pages} {196} (\bibinfo {year} {1986})}\BibitemShut
  {NoStop}%
\bibitem [{\citenamefont {Jaeckel}\ and\ \citenamefont
  {Ringwald}(2010)}]{Jaeckel:2010ni}%
  \BibitemOpen
  \bibfield  {author} {\bibinfo {author} {\bibfnamefont {J.}~\bibnamefont
  {Jaeckel}}\ and\ \bibinfo {author} {\bibfnamefont {A.}~\bibnamefont
  {Ringwald}},\ }\href {\doibase 10.1146/annurev.nucl.012809.104433} {\bibfield
   {journal} {\bibinfo  {journal} {Ann. Rev. Nucl. Part. Sci.}\ }\textbf
  {\bibinfo {volume} {60}},\ \bibinfo {pages} {405} (\bibinfo {year} {2010})},\
  \Eprint {http://arxiv.org/abs/1002.0329} {arXiv:1002.0329 [hep-ph]}
  \BibitemShut {NoStop}%
\bibitem [{\citenamefont {Caputo}\ \emph {et~al.}(2021)\citenamefont {Caputo},
  \citenamefont {Millar}, \citenamefont {O'Hare},\ and\ \citenamefont
  {Vitagliano}}]{Caputo:2021eaa}%
  \BibitemOpen
  \bibfield  {author} {\bibinfo {author} {\bibfnamefont {A.}~\bibnamefont
  {Caputo}}, \bibinfo {author} {\bibfnamefont {A.~J.}\ \bibnamefont {Millar}},
  \bibinfo {author} {\bibfnamefont {C.~A.~J.}\ \bibnamefont {O'Hare}}, \ and\
  \bibinfo {author} {\bibfnamefont {E.}~\bibnamefont {Vitagliano}},\ }\href
  {\doibase 10.1103/PhysRevD.104.095029} {\bibfield  {journal} {\bibinfo
  {journal} {Phys. Rev. D}\ }\textbf {\bibinfo {volume} {104}},\ \bibinfo
  {pages} {095029} (\bibinfo {year} {2021})},\ \Eprint
  {http://arxiv.org/abs/2105.04565} {arXiv:2105.04565 [hep-ph]} \BibitemShut
  {NoStop}%
\bibitem [{\citenamefont {Okun}(1982)}]{Okun:1982xi}%
  \BibitemOpen
  \bibfield  {author} {\bibinfo {author} {\bibfnamefont {L.~B.}\ \bibnamefont
  {Okun}},\ }\href@noop {} {\bibfield  {journal} {\bibinfo  {journal} {Sov.
  Phys. JETP}\ }\textbf {\bibinfo {volume} {56}},\ \bibinfo {pages} {502}
  (\bibinfo {year} {1982})}\BibitemShut {NoStop}%
\bibitem [{\citenamefont {Van~Bibber}\ \emph {et~al.}(1987)\citenamefont
  {Van~Bibber}, \citenamefont {Dagdeviren}, \citenamefont {Koonin},
  \citenamefont {Kerman},\ and\ \citenamefont {Nelson}}]{VanBibber:1987rq}%
  \BibitemOpen
  \bibfield  {author} {\bibinfo {author} {\bibfnamefont {K.}~\bibnamefont
  {Van~Bibber}}, \bibinfo {author} {\bibfnamefont {N.~R.}\ \bibnamefont
  {Dagdeviren}}, \bibinfo {author} {\bibfnamefont {S.~E.}\ \bibnamefont
  {Koonin}}, \bibinfo {author} {\bibfnamefont {A.}~\bibnamefont {Kerman}}, \
  and\ \bibinfo {author} {\bibfnamefont {H.~N.}\ \bibnamefont {Nelson}},\
  }\href {\doibase 10.1103/PhysRevLett.59.759} {\bibfield  {journal} {\bibinfo
  {journal} {Phys. Rev. Lett.}\ }\textbf {\bibinfo {volume} {59}},\ \bibinfo
  {pages} {759} (\bibinfo {year} {1987})}\BibitemShut {NoStop}%
\bibitem [{\citenamefont {Hoogeveen}\ and\ \citenamefont
  {Ziegenhagen}(1991)}]{Hoogeveen:1990vq}%
  \BibitemOpen
  \bibfield  {author} {\bibinfo {author} {\bibfnamefont {F.}~\bibnamefont
  {Hoogeveen}}\ and\ \bibinfo {author} {\bibfnamefont {T.}~\bibnamefont
  {Ziegenhagen}},\ }\href {\doibase 10.1016/0550-3213(91)90528-6} {\bibfield
  {journal} {\bibinfo  {journal} {Nucl. Phys. B}\ }\textbf {\bibinfo {volume}
  {358}},\ \bibinfo {pages} {3} (\bibinfo {year} {1991})}\BibitemShut {NoStop}%
\bibitem [{\citenamefont {Ehret}\ \emph {et~al.}(2009)\citenamefont {Ehret}
  \emph {et~al.}}]{Ehret:2009sq}%
  \BibitemOpen
  \bibfield  {author} {\bibinfo {author} {\bibfnamefont {K.}~\bibnamefont
  {Ehret}} \emph {et~al.} (\bibinfo {collaboration} {ALPS}),\ }\href {\doibase
  10.1016/j.nima.2009.10.102} {\bibfield  {journal} {\bibinfo  {journal} {Nucl.
  Instrum. Meth. A}\ }\textbf {\bibinfo {volume} {612}},\ \bibinfo {pages} {83}
  (\bibinfo {year} {2009})},\ \Eprint {http://arxiv.org/abs/0905.4159}
  {arXiv:0905.4159 [physics.ins-det]} \BibitemShut {NoStop}%
\bibitem [{\citenamefont {Betz}\ \emph {et~al.}(2013)\citenamefont {Betz},
  \citenamefont {Caspers}, \citenamefont {Gasior}, \citenamefont {Thumm},\ and\
  \citenamefont {Rieger}}]{Betz:2013dza}%
  \BibitemOpen
  \bibfield  {author} {\bibinfo {author} {\bibfnamefont {M.}~\bibnamefont
  {Betz}}, \bibinfo {author} {\bibfnamefont {F.}~\bibnamefont {Caspers}},
  \bibinfo {author} {\bibfnamefont {M.}~\bibnamefont {Gasior}}, \bibinfo
  {author} {\bibfnamefont {M.}~\bibnamefont {Thumm}}, \ and\ \bibinfo {author}
  {\bibfnamefont {S.~W.}\ \bibnamefont {Rieger}},\ }\href {\doibase
  10.1103/PhysRevD.88.075014} {\bibfield  {journal} {\bibinfo  {journal} {Phys.
  Rev. D}\ }\textbf {\bibinfo {volume} {88}},\ \bibinfo {pages} {075014}
  (\bibinfo {year} {2013})},\ \Eprint {http://arxiv.org/abs/1310.8098}
  {arXiv:1310.8098 [physics.ins-det]} \BibitemShut {NoStop}%
\bibitem [{\citenamefont {Parker}\ \emph {et~al.}(2013)\citenamefont {Parker},
  \citenamefont {Hartnett}, \citenamefont {Povey},\ and\ \citenamefont
  {Tobar}}]{Parker:2013fxa}%
  \BibitemOpen
  \bibfield  {author} {\bibinfo {author} {\bibfnamefont {S.~R.}\ \bibnamefont
  {Parker}}, \bibinfo {author} {\bibfnamefont {J.~G.}\ \bibnamefont
  {Hartnett}}, \bibinfo {author} {\bibfnamefont {R.~G.}\ \bibnamefont {Povey}},
  \ and\ \bibinfo {author} {\bibfnamefont {M.~E.}\ \bibnamefont {Tobar}},\
  }\href {\doibase 10.1103/PhysRevD.88.112004} {\bibfield  {journal} {\bibinfo
  {journal} {Phys. Rev. D}\ }\textbf {\bibinfo {volume} {88}},\ \bibinfo
  {pages} {112004} (\bibinfo {year} {2013})},\ \Eprint
  {http://arxiv.org/abs/1410.5244} {arXiv:1410.5244 [hep-ex]} \BibitemShut
  {NoStop}%
\bibitem [{\citenamefont {Jaeckel}\ and\ \citenamefont
  {Ringwald}(2008)}]{Jaeckel_PhysLettB_2008}%
  \BibitemOpen
  \bibfield  {author} {\bibinfo {author} {\bibfnamefont {J.}~\bibnamefont
  {Jaeckel}}\ and\ \bibinfo {author} {\bibfnamefont {A.}~\bibnamefont
  {Ringwald}},\ }\href {\doibase
  http://dx.doi.org/10.1016/j.physletb.2007.11.071} {\bibfield  {journal}
  {\bibinfo  {journal} {Phys. Lett. B}\ }\textbf {\bibinfo {volume} {659}},\
  \bibinfo {pages} {509 } (\bibinfo {year} {2008})}\BibitemShut {NoStop}%
\bibitem [{\citenamefont {Graham}\ \emph {et~al.}(2014)\citenamefont {Graham},
  \citenamefont {Mardon}, \citenamefont {Rajendran},\ and\ \citenamefont
  {Zhao}}]{Graham_PRD_2014}%
  \BibitemOpen
  \bibfield  {author} {\bibinfo {author} {\bibfnamefont {P.~W.}\ \bibnamefont
  {Graham}}, \bibinfo {author} {\bibfnamefont {J.}~\bibnamefont {Mardon}},
  \bibinfo {author} {\bibfnamefont {S.}~\bibnamefont {Rajendran}}, \ and\
  \bibinfo {author} {\bibfnamefont {Y.}~\bibnamefont {Zhao}},\ }\href {\doibase
  10.1103/PhysRevD.90.075017} {\bibfield  {journal} {\bibinfo  {journal} {Phys.
  Rev. D}\ }\textbf {\bibinfo {volume} {90}},\ \bibinfo {pages} {075017}
  (\bibinfo {year} {2014})}\BibitemShut {NoStop}%
\bibitem [{\citenamefont {Padamsee}(2014)}]{Padamsee_Ann_Rev_Nucl_2014}%
  \BibitemOpen
  \bibfield  {author} {\bibinfo {author} {\bibfnamefont {H.~S.}\ \bibnamefont
  {Padamsee}},\ }\href {\doibase 10.1146/annurev-nucl-102313-025612} {\bibfield
   {journal} {\bibinfo  {journal} {Annual Review of Nuclear and Particle
  Science}\ }\textbf {\bibinfo {volume} {64}},\ \bibinfo {pages} {175}
  (\bibinfo {year} {2014})}\BibitemShut {NoStop}%
\bibitem [{\citenamefont {Dicke}(1946)}]{Dicke:1946glx}%
  \BibitemOpen
  \bibfield  {author} {\bibinfo {author} {\bibfnamefont {R.~H.}\ \bibnamefont
  {Dicke}},\ }\href {\doibase 10.1063/1.1770483} {\bibfield  {journal}
  {\bibinfo  {journal} {Rev. Sci. Instrum.}\ }\textbf {\bibinfo {volume}
  {17}},\ \bibinfo {pages} {268} (\bibinfo {year} {1946})}\BibitemShut
  {NoStop}%
\bibitem [{\citenamefont {Pischalnikov}\ \emph {et~al.}(2015)\citenamefont
  {Pischalnikov}, \citenamefont {Borissov}, \citenamefont {Gonin},
  \citenamefont {Holzbauer}, \citenamefont {Khabiboulline}, \citenamefont
  {Schappert}, \citenamefont {Smith},\ and\ \citenamefont
  {Yun}}]{Pischalnikov_IPAC2015}%
  \BibitemOpen
  \bibfield  {author} {\bibinfo {author} {\bibfnamefont {Y.}~\bibnamefont
  {Pischalnikov}}, \bibinfo {author} {\bibfnamefont {E.}~\bibnamefont
  {Borissov}}, \bibinfo {author} {\bibfnamefont {I.}~\bibnamefont {Gonin}},
  \bibinfo {author} {\bibfnamefont {J.}~\bibnamefont {Holzbauer}}, \bibinfo
  {author} {\bibfnamefont {T.}~\bibnamefont {Khabiboulline}}, \bibinfo {author}
  {\bibfnamefont {W.}~\bibnamefont {Schappert}}, \bibinfo {author}
  {\bibfnamefont {S.}~\bibnamefont {Smith}}, \ and\ \bibinfo {author}
  {\bibfnamefont {J.}~\bibnamefont {Yun}},\ }in\ \href
  {http://accelconf.web.cern.ch/IPAC2015/papers/wepty035.pdf} {\emph {\bibinfo
  {booktitle} {Proceedings of the 6th International Particle Accelerator
  Conference}}},\ \bibinfo {series and number} {\bibinfo {number} {WEPTY035}}\
  (\bibinfo {year} {2015})\BibitemShut {NoStop}%
\bibitem [{\citenamefont {Pischalnikov}\ \emph
  {et~al.}(2019{\natexlab{a}})\citenamefont {Pischalnikov}, \citenamefont
  {Bice}, \citenamefont {Grassellino}, \citenamefont {Khabiboulline},
  \citenamefont {Melnychuk}, \citenamefont {Pilipenko}, \citenamefont {Posen},
  \citenamefont {Pronichev},\ and\ \citenamefont
  {Romanenko}}]{Pischalnikov_SRF2019}%
  \BibitemOpen
  \bibfield  {author} {\bibinfo {author} {\bibfnamefont {Y.}~\bibnamefont
  {Pischalnikov}}, \bibinfo {author} {\bibfnamefont {D.}~\bibnamefont {Bice}},
  \bibinfo {author} {\bibfnamefont {A.}~\bibnamefont {Grassellino}}, \bibinfo
  {author} {\bibfnamefont {T.}~\bibnamefont {Khabiboulline}}, \bibinfo {author}
  {\bibfnamefont {O.}~\bibnamefont {Melnychuk}}, \bibinfo {author}
  {\bibfnamefont {R.}~\bibnamefont {Pilipenko}}, \bibinfo {author}
  {\bibfnamefont {S.}~\bibnamefont {Posen}}, \bibinfo {author} {\bibfnamefont
  {O.}~\bibnamefont {Pronichev}}, \ and\ \bibinfo {author} {\bibfnamefont
  {A.}~\bibnamefont {Romanenko}},\ }in\ \href
  {https://accelconf.web.cern.ch/srf2019/papers/tup085.pdf} {\emph {\bibinfo
  {booktitle} {Proceedings of the 19th International Conference on RF
  Superconductivity}}},\ \bibinfo {series and number} {\bibinfo {number}
  {TUP085}}\ (\bibinfo {year} {2019})\BibitemShut {NoStop}%
\bibitem [{\citenamefont {Melnychuk}\ \emph {et~al.}(2014)\citenamefont
  {Melnychuk}, \citenamefont {Grassellino},\ and\ \citenamefont
  {Romanenko}}]{Melnychuk_RSI_2014}%
  \BibitemOpen
  \bibfield  {author} {\bibinfo {author} {\bibfnamefont {O.}~\bibnamefont
  {Melnychuk}}, \bibinfo {author} {\bibfnamefont {A.}~\bibnamefont
  {Grassellino}}, \ and\ \bibinfo {author} {\bibfnamefont {A.}~\bibnamefont
  {Romanenko}},\ }\href {\doibase 10.1063/1.4903868} {\bibfield  {journal}
  {\bibinfo  {journal} {Rev. Sci. Instrum.}\ }\textbf {\bibinfo {volume}
  {85}},\ \bibinfo {pages} {124705} (\bibinfo {year} {2014})}\BibitemShut
  {NoStop}%
\bibitem [{\citenamefont {Pischalnikov}\ \emph
  {et~al.}(2019{\natexlab{b}})\citenamefont {Pischalnikov}, \citenamefont
  {Bice}, \citenamefont {Grassellino}, \citenamefont {Khabiboulline},
  \citenamefont {Melnychuk}, \citenamefont {Pilipenko}, \citenamefont {Posen},
  \citenamefont {Pronichev},\ and\ \citenamefont
  {Romanenko}}]{Pischalnikov:2019iyu}%
  \BibitemOpen
  \bibfield  {author} {\bibinfo {author} {\bibfnamefont {Y.}~\bibnamefont
  {Pischalnikov}}, \bibinfo {author} {\bibfnamefont {D.}~\bibnamefont {Bice}},
  \bibinfo {author} {\bibfnamefont {A.}~\bibnamefont {Grassellino}}, \bibinfo
  {author} {\bibfnamefont {T.}~\bibnamefont {Khabiboulline}}, \bibinfo {author}
  {\bibfnamefont {O.}~\bibnamefont {Melnychuk}}, \bibinfo {author}
  {\bibfnamefont {R.}~\bibnamefont {Pilipenko}}, \bibinfo {author}
  {\bibfnamefont {S.}~\bibnamefont {Posen}}, \bibinfo {author} {\bibfnamefont
  {O.}~\bibnamefont {Pronichev}}, \ and\ \bibinfo {author} {\bibfnamefont
  {A.}~\bibnamefont {Romanenko}}\ }(\bibinfo {year} {2019})\BibitemShut
  {NoStop}%
\bibitem [{\citenamefont {Mirizzi}\ \emph {et~al.}(2009)\citenamefont
  {Mirizzi}, \citenamefont {Redondo},\ and\ \citenamefont
  {Sigl}}]{Mirizzi:2009iz}%
  \BibitemOpen
  \bibfield  {author} {\bibinfo {author} {\bibfnamefont {A.}~\bibnamefont
  {Mirizzi}}, \bibinfo {author} {\bibfnamefont {J.}~\bibnamefont {Redondo}}, \
  and\ \bibinfo {author} {\bibfnamefont {G.}~\bibnamefont {Sigl}},\ }\href
  {\doibase 10.1088/1475-7516/2009/03/026} {\bibfield  {journal} {\bibinfo
  {journal} {JCAP}\ }\textbf {\bibinfo {volume} {03}},\ \bibinfo {pages} {026}
  (\bibinfo {year} {2009})},\ \Eprint {http://arxiv.org/abs/0901.0014}
  {arXiv:0901.0014 [hep-ph]} \BibitemShut {NoStop}%
\bibitem [{\citenamefont {Caputo}\ \emph {et~al.}(2020)\citenamefont {Caputo},
  \citenamefont {Liu}, \citenamefont {Mishra-Sharma},\ and\ \citenamefont
  {Ruderman}}]{Caputo:2020bdy}%
  \BibitemOpen
  \bibfield  {author} {\bibinfo {author} {\bibfnamefont {A.}~\bibnamefont
  {Caputo}}, \bibinfo {author} {\bibfnamefont {H.}~\bibnamefont {Liu}},
  \bibinfo {author} {\bibfnamefont {S.}~\bibnamefont {Mishra-Sharma}}, \ and\
  \bibinfo {author} {\bibfnamefont {J.~T.}\ \bibnamefont {Ruderman}},\ }\href
  {\doibase 10.1103/PhysRevLett.125.221303} {\bibfield  {journal} {\bibinfo
  {journal} {Phys. Rev. Lett.}\ }\textbf {\bibinfo {volume} {125}},\ \bibinfo
  {pages} {221303} (\bibinfo {year} {2020})},\ \Eprint
  {http://arxiv.org/abs/2002.05165} {arXiv:2002.05165 [astro-ph.CO]}
  \BibitemShut {NoStop}%
\bibitem [{\citenamefont {Garcia}\ \emph {et~al.}(2020)\citenamefont {Garcia},
  \citenamefont {Bondarenko}, \citenamefont {Ploeckinger}, \citenamefont
  {Pradler},\ and\ \citenamefont {Sokolenko}}]{Garcia:2020qrp}%
  \BibitemOpen
  \bibfield  {author} {\bibinfo {author} {\bibfnamefont {A.~A.}\ \bibnamefont
  {Garcia}}, \bibinfo {author} {\bibfnamefont {K.}~\bibnamefont {Bondarenko}},
  \bibinfo {author} {\bibfnamefont {S.}~\bibnamefont {Ploeckinger}}, \bibinfo
  {author} {\bibfnamefont {J.}~\bibnamefont {Pradler}}, \ and\ \bibinfo
  {author} {\bibfnamefont {A.}~\bibnamefont {Sokolenko}},\ }\href {\doibase
  10.1088/1475-7516/2020/10/011} {\bibfield  {journal} {\bibinfo  {journal}
  {JCAP}\ }\textbf {\bibinfo {volume} {10}},\ \bibinfo {pages} {011} (\bibinfo
  {year} {2020})},\ \Eprint {http://arxiv.org/abs/2003.10465} {arXiv:2003.10465
  [astro-ph.CO]} \BibitemShut {NoStop}%
\bibitem [{\citenamefont {Williams}\ \emph {et~al.}(1971)\citenamefont
  {Williams}, \citenamefont {Faller},\ and\ \citenamefont
  {Hill}}]{Williams:1971ms}%
  \BibitemOpen
  \bibfield  {author} {\bibinfo {author} {\bibfnamefont {E.~R.}\ \bibnamefont
  {Williams}}, \bibinfo {author} {\bibfnamefont {J.~E.}\ \bibnamefont
  {Faller}}, \ and\ \bibinfo {author} {\bibfnamefont {H.~A.}\ \bibnamefont
  {Hill}},\ }\href {\doibase 10.1103/PhysRevLett.26.721} {\bibfield  {journal}
  {\bibinfo  {journal} {Phys. Rev. Lett.}\ }\textbf {\bibinfo {volume} {26}},\
  \bibinfo {pages} {721} (\bibinfo {year} {1971})}\BibitemShut {NoStop}%
\bibitem [{\citenamefont {Abel}\ \emph {et~al.}(2008)\citenamefont {Abel},
  \citenamefont {Goodsell}, \citenamefont {Jaeckel}, \citenamefont {Khoze},\
  and\ \citenamefont {Ringwald}}]{Abel:2008ai}%
  \BibitemOpen
  \bibfield  {author} {\bibinfo {author} {\bibfnamefont {S.~A.}\ \bibnamefont
  {Abel}}, \bibinfo {author} {\bibfnamefont {M.~D.}\ \bibnamefont {Goodsell}},
  \bibinfo {author} {\bibfnamefont {J.}~\bibnamefont {Jaeckel}}, \bibinfo
  {author} {\bibfnamefont {V.~V.}\ \bibnamefont {Khoze}}, \ and\ \bibinfo
  {author} {\bibfnamefont {A.}~\bibnamefont {Ringwald}},\ }\href {\doibase
  10.1088/1126-6708/2008/07/124} {\bibfield  {journal} {\bibinfo  {journal}
  {JHEP}\ }\textbf {\bibinfo {volume} {07}},\ \bibinfo {pages} {124} (\bibinfo
  {year} {2008})},\ \Eprint {http://arxiv.org/abs/0803.1449} {arXiv:0803.1449
  [hep-ph]} \BibitemShut {NoStop}%
\bibitem [{\citenamefont {Workman}\ \emph {et~al.}(2022)\citenamefont {Workman}
  \emph {et~al.}}]{ParticleDataGroup:2022pth}%
  \BibitemOpen
  \bibfield  {author} {\bibinfo {author} {\bibfnamefont {R.~L.}\ \bibnamefont
  {Workman}} \emph {et~al.} (\bibinfo {collaboration} {Particle Data Group}),\
  }\href {\doibase 10.1093/ptep/ptac097} {\bibfield  {journal} {\bibinfo
  {journal} {PTEP}\ }\textbf {\bibinfo {volume} {2022}},\ \bibinfo {pages}
  {083C01} (\bibinfo {year} {2022})}\BibitemShut {NoStop}%
\bibitem [{\citenamefont {Berlin}\ \emph
  {et~al.}(2022{\natexlab{a}})\citenamefont {Berlin}, \citenamefont {Dror},
  \citenamefont {Gan},\ and\ \citenamefont {Ruderman}}]{Berlin:2022hmt}%
  \BibitemOpen
  \bibfield  {author} {\bibinfo {author} {\bibfnamefont {A.}~\bibnamefont
  {Berlin}}, \bibinfo {author} {\bibfnamefont {J.~A.}\ \bibnamefont {Dror}},
  \bibinfo {author} {\bibfnamefont {X.}~\bibnamefont {Gan}}, \ and\ \bibinfo
  {author} {\bibfnamefont {J.~T.}\ \bibnamefont {Ruderman}},\ }\href@noop {} {\
   (\bibinfo {year} {2022}{\natexlab{a}})},\ \Eprint
  {http://arxiv.org/abs/2211.05139} {arXiv:2211.05139 [hep-ph]} \BibitemShut
  {NoStop}%
\bibitem [{\citenamefont {Workman}\ and\ \citenamefont
  {Others}(2022)}]{Workman:2022ynf}%
  \BibitemOpen
  \bibfield  {author} {\bibinfo {author} {\bibfnamefont {R.~L.}\ \bibnamefont
  {Workman}}\ and\ \bibinfo {author} {\bibnamefont {Others}} (\bibinfo
  {collaboration} {Particle Data Group}),\ }\href {\doibase
  10.1093/ptep/ptac097} {\bibfield  {journal} {\bibinfo  {journal} {PTEP}\
  }\textbf {\bibinfo {volume} {2022}},\ \bibinfo {pages} {083C01} (\bibinfo
  {year} {2022})}\BibitemShut {NoStop}%
\bibitem [{\citenamefont {Berlin}\ and\ \citenamefont
  {Hook}(2020)}]{Berlin:2020pey}%
  \BibitemOpen
  \bibfield  {author} {\bibinfo {author} {\bibfnamefont {A.}~\bibnamefont
  {Berlin}}\ and\ \bibinfo {author} {\bibfnamefont {A.}~\bibnamefont {Hook}},\
  }\href {\doibase 10.1103/PhysRevD.102.035010} {\bibfield  {journal} {\bibinfo
   {journal} {Phys. Rev. D}\ }\textbf {\bibinfo {volume} {102}},\ \bibinfo
  {pages} {035010} (\bibinfo {year} {2020})},\ \Eprint
  {http://arxiv.org/abs/2001.02679} {arXiv:2001.02679 [hep-ph]} \BibitemShut
  {NoStop}%
\bibitem [{\citenamefont {Ahlers}\ \emph {et~al.}(2007)\citenamefont {Ahlers},
  \citenamefont {Gies}, \citenamefont {Jaeckel},\ and\ \citenamefont
  {Ringwald}}]{Ahlers:2006iz}%
  \BibitemOpen
  \bibfield  {author} {\bibinfo {author} {\bibfnamefont {M.}~\bibnamefont
  {Ahlers}}, \bibinfo {author} {\bibfnamefont {H.}~\bibnamefont {Gies}},
  \bibinfo {author} {\bibfnamefont {J.}~\bibnamefont {Jaeckel}}, \ and\
  \bibinfo {author} {\bibfnamefont {A.}~\bibnamefont {Ringwald}},\ }\href
  {\doibase 10.1103/PhysRevD.75.035011} {\bibfield  {journal} {\bibinfo
  {journal} {Phys. Rev. D}\ }\textbf {\bibinfo {volume} {75}},\ \bibinfo
  {pages} {035011} (\bibinfo {year} {2007})},\ \Eprint
  {http://arxiv.org/abs/hep-ph/0612098} {arXiv:hep-ph/0612098} \BibitemShut
  {NoStop}%
\bibitem [{\citenamefont {Della~Valle}\ \emph {et~al.}(2014)\citenamefont
  {Della~Valle}, \citenamefont {Milotti}, \citenamefont {Ejlli}, \citenamefont
  {Messineo}, \citenamefont {Piemontese}, \citenamefont {Zavattini},
  \citenamefont {Gastaldi}, \citenamefont {Pengo},\ and\ \citenamefont
  {Ruoso}}]{DellaValle:2014xoa}%
  \BibitemOpen
  \bibfield  {author} {\bibinfo {author} {\bibfnamefont {F.}~\bibnamefont
  {Della~Valle}}, \bibinfo {author} {\bibfnamefont {E.}~\bibnamefont
  {Milotti}}, \bibinfo {author} {\bibfnamefont {A.}~\bibnamefont {Ejlli}},
  \bibinfo {author} {\bibfnamefont {G.}~\bibnamefont {Messineo}}, \bibinfo
  {author} {\bibfnamefont {L.}~\bibnamefont {Piemontese}}, \bibinfo {author}
  {\bibfnamefont {G.}~\bibnamefont {Zavattini}}, \bibinfo {author}
  {\bibfnamefont {U.}~\bibnamefont {Gastaldi}}, \bibinfo {author}
  {\bibfnamefont {R.}~\bibnamefont {Pengo}}, \ and\ \bibinfo {author}
  {\bibfnamefont {G.}~\bibnamefont {Ruoso}},\ }\href {\doibase
  10.1103/PhysRevD.90.092003} {\bibfield  {journal} {\bibinfo  {journal} {Phys.
  Rev. D}\ }\textbf {\bibinfo {volume} {90}},\ \bibinfo {pages} {092003}
  (\bibinfo {year} {2014})},\ \Eprint {http://arxiv.org/abs/1406.6518}
  {arXiv:1406.6518 [quant-ph]} \BibitemShut {NoStop}%
\bibitem [{\citenamefont {Berlin}\ \emph
  {et~al.}(2022{\natexlab{b}})\citenamefont {Berlin} \emph
  {et~al.}}]{Berlin:2022hfx}%
  \BibitemOpen
  \bibfield  {author} {\bibinfo {author} {\bibfnamefont {A.}~\bibnamefont
  {Berlin}} \emph {et~al.},\ }\href@noop {} {\  (\bibinfo {year}
  {2022}{\natexlab{b}})},\ \Eprint {http://arxiv.org/abs/2203.12714}
  {arXiv:2203.12714 [hep-ph]} \BibitemShut {NoStop}%
\end{thebibliography}%


\appendix

\section{Signal Strength Calculation}

Here we present the calculation of the signal power in the receiver cavity using the formalism in Ref.~\cite{Graham_PRD_2014}. 
For simplicity, we assume two identical cavities and focus on a particular cavity mode with frequency $\omega$ and with a field $\vec E (\vec x, t) = \vec E_\mathrm{cav}(\vec x) e^{i\omega t}$.
The dark photon field sourced by the emitter cavity is approximately
\begin{equation}
    \vec E^\prime(\vec r, t)\simeq-\epsilon \, m_{\gamma^\prime}^2 \int_{V_\mathrm{emitter}} \hspace{-0.4cm} d^3 x ~ \frac {\vec E_\mathrm{cav}(\vec x)} {4\pi |\vec r-\vec x|} ~ e^{i(\omega t -k|\vec r-\vec x|)},
    \label{eq:darkfield}
\end{equation}
where $V_\mathrm{emitter}$ is the emitter cavity volume, and $k^2=\omega^2-m_{\gamma^\prime}^2$. 
For $m_{\gamma^\prime}>\omega$, $k=-i\sqrt{-\omega^2+m_{\gamma^\prime}^2}$. 

The induced effective current density is
\begin{equation}
    \vec \jmath(\vec r) e^{i\omega t}=-\frac {i \epsilon} {\omega}\left(m_{\gamma^\prime}^2\vec E^\prime-\vec\nabla(\vec \nabla\cdot \vec E^\prime)\right).
\end{equation}
In the limit where the emitter and receiver frequency match, the observable signal in the receiver cavity reads
\begin{equation}
    \vec E_\mathrm{receiver}(\vec r, t)=-\frac {Q_{\rm rec}} \omega \left[\frac {\int d^3x \vec E^*_\mathrm{cav}(\vec x)\cdot \vec\jmath(\vec x)} {\int d^3x|\vec E_\mathrm{cav}(\vec x)|^2}\right] \vec E_\mathrm{cav}(\vec r) e^{i\omega t},
    \label{eq:signalE}
\end{equation}
where the $\int d^3x$ integrates over the receiver cavity volume, $\vec E_{\rm cav}(\vec x)$ represents the electric field strength of a given receiver cavity mode. 
It is convenient to identify the (normalized) quantity in the square brackets above as a coupling or form factor

\begin{equation}
    |G|^2\equiv \frac 1 {\epsilon^4 }\left(\frac {\omega} {m_{\gamma^\prime}}\right)^4 \left[\frac {\int d^3x \vec E^*_\mathrm{cav}(\vec x)\cdot \vec\jmath(\vec x)} {\omega \int d^3x|\vec E_\mathrm{cav}(\vec x)|^2}\right]^2,
\end{equation}
where the effective current $\vec\jmath$ depends on receiver  positioning $\vec r$,  dark photon mass $m_{\gamma^\prime}$, and $\epsilon$. With our positioning and dark photon mass around $\omega$, the longitudinal component dominates. This drives us considering the prefactor/normalization of $|G|^2$, which becomes independent of $Q_{\rm rec}$, $\epsilon$ and $m_{\gamma^\prime}$ to leading order.

In our experiments, we need to consider the frequency spread from both the emitter and the receiver, the frequency drift over time, and frequency jittering due to other mechanical effects such as bubble collisions. These effects can be {\it conservatively}\footnote{Jittering, for instance, can be modeled as a reduction of integration time, which would be a lesser suppression than what we use here.} modeled as a mismatch between receiver frequency and emitter frequency, effectively reducing the form factor $|G|^2$ by 

\begin{equation}
|G|^2\rightarrow \frac {\omega^2} {\omega^2+4 \delta_\omega^2 Q^2_{\rm rec}}|G|^2\,,
\end{equation}
where $\delta_\omega$ represents a typical mismatch in (angular) frequency. For our run that determines the new results, the frequency mismatch is conservatively assumed to be $7.8$~Hz (from frequency drift and jittering). With the resonant frequency of 1.3~GHz, and receiver cavity intrinsic $Q$ factor of  $3\times 10^{10}$, jittering causes a suppression of the signal power by a factor of $7.7\times 10^{-6}$.

\end{document}